\documentclass[a4paper,12pt]{article}
\usepackage{amsmath}
\usepackage{amsfonts}
\usepackage{amssymb}
\usepackage[dvipdfm]{graphicx}
\usepackage{amsthm}
\usepackage{mathrsfs}
\usepackage{mathtools}
\usepackage{mathrsfs}
\usepackage{amsrefs}
\usepackage{color}
\DeclarePairedDelimiter{\abs}{\lvert}{\rvert}
\numberwithin{equation}{section}

\theoremstyle{definition}
\newtheorem{Def}{Definition}[section]
\newtheorem{Prop}[Def]{Proposition}
\newtheorem{Th}[Def]{Theorem}
\newtheorem{Lem}[Def]{Lemma}

\newtheorem{Con}[Def]{Condition}

\title{\Large A Rellich type theorem for the generalized oscillator \\  }
\author {T. Tagawa \\ }
\date{}

\begin{document}
\maketitle
\begin{abstract}
For the generalized oscillator, we prove a Rellich type theorem, or characterize the order of growth of eigenfunctions. The proofs are given by an extensive use of commutator arguments invented recently by Ito and Skibsted. These arguments are simple and elementary and do not employ energy cut-offs or microlocal analysis.
\end{abstract}
\section{Introduction}
\subsection{Setting}
For any fixed $a,b>0$ we consider the generalized oscillator
\[H = -\frac{1}{2}\Delta + \frac{a}{2} \abs{x}^{2b} + V;\  -\Delta =p_j\delta^{jk}p_k,\ p_j = -i\partial_{x_j}, \]
on the Hilbert space $\mathcal{H}=L^{2}(\mathbb{R}^d),\ d \geq 1$. Here $\delta^{jk}$ is the Kronecker delta, we use the Einstein summation convention (throughout the paper we will use this notation) and $V$ is the real-valued function that may grow slightly slower than $ \frac{a}{2} \abs{x}^{2b}$.

Choose $\chi \in C^{\infty}(\mathbb{R})$ such that 
\begin{equation}
\label{chi}
\chi(t) =
  \begin{cases}
  1 & \text{for $t$ $\leq$ 1,}\\
  0 & \text{for $t$ $\geq$ 2,}
  \end{cases}
\  \chi^{\prime} \leq 0,
\end{equation}
and set $r \in C^{\infty}(\mathbb{R}^{d})$ as 
\begin{equation}
\label{r}
r(x) = \chi (\abs x) + \abs x(1-\chi(\abs x)).
\end{equation}
\begin{Con}
\label{Con}
The perturbation $V$ is a real-valued function. Moreover, there exists a splititng by real-valued functions:
\begin{equation*}
V = V_1 +V_2;\  V_1 \in C^{1}(\mathbb{R}^d;\mathbb{R}),
\end{equation*}
such that for some $\mu, C >0$ the following bounds hold globally on $\mathbb{R}^{d}$:
\begin{equation}
\label{V}
\abs{V_1} \leq Cr^{2b-\mu},\ \abs{ \partial_{r}V_1} \leq Cr^{2b-1-\mu},\ \abs{V_2} \leq Cr^{b-1-\mu }.
\end{equation} 
Here $\partial_{r} = (\partial_{i} r)\partial_{i} $ denotes the radial differential operator.
\end{Con}

We introduce the weighted Hilbert space $\mathcal{H}_{s}$ for $s \in \mathbb{R}$ by
\begin{equation*}
\mathcal{H}_{s}=r^{-s}\mathcal{H}.
\end{equation*}
We also denote the locally $L^{2}$-space by
\begin{equation*}
\mathcal{H}_\mathrm{loc}=L^2_\mathrm{loc}(\mathbb{R}^d).
\end{equation*}
\subsection{Results}
Our main results are the following three theorems. In this paper we assume Condition $\ref{Con}$.
\begin{Th}
\label{th1}
Let $\lambda \in \mathbb{R}$. If a function $\phi \in \mathcal{H}_\mathrm{loc}$ satisfies that
\begin{itemize}
\item $(H-\lambda)\phi$ =0 in the distributional sense,
\item there exists an $\alpha > \frac{\sqrt{a}}{1+b}$ such that  $\exp({\alpha \abs {x}^{b+1}}) \phi \in \mathcal{H}$, 
\end{itemize} 
then $\phi =0$ in $\mathbb{R}^{d}$.
\end{Th}
\begin{Th}
\label{th2}
Let $\lambda \in \mathbb{R}$. If a function $\phi \in \mathcal{H}_\mathrm{loc}$ satisfies that
\begin{itemize}
\item $(H-\lambda)\phi$ =0 in the distributional sense,
\item there exists an $\abs{\alpha} < \frac{\sqrt{a}}{1+b}$ such that  $\exp({\alpha \abs {x}^{b+1}}) \phi \in \mathcal{H}$, 
\end{itemize} 
then $\exp({\alpha \abs {x}^{b+1}}) \phi \in \mathcal{H}$ for any $\abs{\alpha} < \frac{\sqrt{a}}{1+b} $.
\end{Th}
To state the third main theorem, we introduce a differential operator 
\begin{equation*}
L = p_{i} {\ell}^{ij} p_{j}\  \mathrm{with} \  {\ell}^{ij} = \delta^{ij} - (\partial_{x_i} r)(\partial_{x_j} r),
\end{equation*}
which may be considered the spherical part of $- \Delta$ on $\{x \in \mathbb{R}^{d} \mid r(x) \geq 2\}$. 
We also use the notation $\langle T \rangle_{\phi}= \langle \phi, T\phi \rangle. $
\begin{Th}
\label{th3}
Let $\lambda \in \mathbb{R}$. If a function $\phi \in \mathcal{H}_\mathrm{loc}$ satisfies that
\begin{itemize}
\item $(H-\lambda)\phi$ =0 in the distributional sense,
\item there exists an $\alpha < -\frac{\sqrt{a}}{1+b}$ such that  $\exp({\alpha \abs {x}^{b+1}}) \phi \in \mathcal{H}$,
\item there exist $C, \rho>0$ such that for any $\chi \in C^{\infty}_{0}(\mathbb{R}^{d})$ the following property holds
\end{itemize} 
\begin{equation*}
\langle p_{i} \chi {\ell}^{ij} p_{j} \rangle _{\phi} \leq C \langle \chi r^{2b-\rho} \rangle_{\phi}, 
\end{equation*}
then $\exp({\alpha \abs {x}^{b+1}}) \phi \in \mathcal{H}$  for any $\alpha < -\frac{\sqrt{a}}{1+b} $. 
\end{Th}
The Schr\"{o}dinger operator corresponding to the usual harmonic oscillator has $L^2$-eigenfunctions of the form $a(x)\exp(- \abs{x}^{2}/2)$ and generalized eigenfunctions of the form $b(x)\exp(\abs{x}^{2}/2)$, where $a(x)$ and $b(x)$ are certain polynomials.
Our main results describe the asymptotic behavior of eigenfunctions for the generalized oscillator.
The first theorem states the non-existence of eigenfunctions that decay stronglier than $\exp(-\frac{\sqrt{a}}{1+b}\abs{x}^{b+1})$.
The second theorem states the non-existence of eigenfunctions with increasing rates between $\exp(-\frac{\sqrt{a}}{1+b}\abs{x}^{b+1})$ and $\exp(\frac{\sqrt{a}}{1+b}\abs{x}^{b+1})$.
The third theorem states the non-existence of eigenfunctions with increasing rates greater than $\exp(\frac{\sqrt{a}}{1+b}\abs{x}^{b+1})$ under the assumption on the angular momentum $L$. 
We note that the constants $\pm\frac{\sqrt{a}}{1+b}$ in the main theorems are optimal. This can be seen from the following.
We set 
\begin{equation*}
\phi_{\pm}(x) = \exp(\pm \frac{ \sqrt{a}}{1+b} r^{b+1} - r^{b}).
\end{equation*} 
Then by taking some appropriate $V_{\pm}$, we have $H\phi_{\pm} =0$.
We also note that the condition on $L$ in the third theorem cannot be removed. 
By giving an example we show this.
Now we consider
\begin{equation*}
H = -\frac{1}{2}\Delta + \frac{1}{2} ({x}^{2}+y^{2}) \ \mathrm{on} \ \mathbb{R}^2.
\end{equation*}
We also set 
\begin{equation*}
\phi(x,y) = \exp(x^{2}+i\sqrt{3}xy -y^{2}).
\end{equation*}
Then we can verify the fact. Of course, this condition holds automatically in one dimension, since $L=0$.

To prove our results we apply the commutator argument invented recently by Ito and Skibsted from \cite{IS}. We are directly motivated by their result in which they studied spectral properties of the Schr\"{o}dinger operator on a manifold with ends. They consider potentials decaying at infinity. In \cite{Itakura}, Itakura proved the non-existence of $B^{*}_{0}$-eigenfunctions for the Schr\"{o}dinger operators with potentials diverging to $-\infty$ at infinity by using the method of the commutator argument. However, they do not consider the Schr\"{o}dinger operator with growing potentials, which are considered in this paper.

In case $b=0$, there are an extensive amount of literature on eigenvalue problems, e.g. \cite{Agmon}, \cite{FRH}, \cite{FRH2O}, \cite{Ho}, \cite{IJ}, \cite{IS}.
As for the case $a,b >0$, Simon studied for the asymptotic behavior of the $L^{2}$-eigenfunctions under a smooth potential in \cite{SB}, but he does not consider the generalized eigenfunctions. Recently Steinerberger discussed a sharp pointwise Agmon estimate in \cite{SS}.
We also mention a result \cite{IM} by Isozaki and Morioka that studies Rellich's theorem for the discrete Schr\"{o}dinger operator.

This paper is organized as follows. In Section 2, 
we first discuss the self-adjoint realization of $H$.
Next, we introduce the conjugate operator $A$. 
Finally, we introduce commutators with weight inside, which play important roles in the proofs of the main results. 
In Section 3, we prove the main theorems.
\section{Preliminaries}
In this section we prepare some tools to prove our theorems.
\subsection{Self-adjoint realization of $H$}
\begin{Lem}
\label{realizaiton}
The operator $H$ is essentially self-adjoint on $C^{\infty}_{0}(\mathbb{R}^d)$
\end{Lem}
\begin{proof}
There exisits some $\gamma>0$ such that  $\frac{a}{2} \abs{x}^{2b} + V + \gamma \geq 0$ on $\mathbb{R}^{d}$. Therefore $H+\gamma$ is the Schr\"{o}dinger operator with positive potential. Hence by \cite[Theorem X.28]{RS}, $H+\gamma$ is essentially self-adjoint on $C^{\infty}_{0}(\mathbb{R}^d)$. This implies the assertion.
\end{proof}
By Lemma $\ref{realizaiton}$ we denote the self-adjoint extension by the same symbol $H$ throughout this paper.

We set 
\begin{equation*} 
H^{2}_{\mathrm{comp}}(\mathbb{R}^d) = \{\psi \in H^2(\mathbb{R}^d) \mid \mathrm{supp}\ \psi\ \mathrm{is}\ \mathrm{compact} \}.
\end{equation*}
\begin{Lem}\label{domain}
The following inclusion relations hold.
\begin{equation*}
H^{2}_{\mathrm{comp}}(\mathbb{R}^d) \subset \mathcal{D}(H) \subset H^1(\mathbb{R}^d).
\end{equation*}
\end{Lem}
\begin{proof}
First, we prove $H^{2}_{\mathrm{comp}}(\mathbb{R}^d) \subset \mathcal{D}(H). $ Let $\psi \in H^{2}_{\mathrm{comp}}(\mathbb{R}^d)$ and $\mathrm{supp}$ $\psi = K$, and set
\begin{equation*}
K_1 = \left\{x \in \mathbb{R}^d \mid \inf_{y \in K} \abs{x-y} \leq 1 \right\}.
\end{equation*} 
Then there exists $\{\psi_n\} \subset C^{\infty}_{0}(\mathbb{R}^d) $ such that 
\begin{equation*}
\mathrm{supp}\ \psi_n  \subset K_1,\  \| \psi_n - \psi\| + \|p^2(\psi_n-\psi)\| \rightarrow 0\   \mathrm{as}\ n \rightarrow \infty. 
\end{equation*}
Hence we can estimate as follows.
\begin{equation*}
\|H \psi_n - H \psi\| + \|\psi_n - \psi\| \leq \frac{1}{2} \|p^2(\psi_n - \psi) \| + C_K \| \psi_n -\psi \| \rightarrow 0\ \mathrm{as}\ n \rightarrow \infty. 
\end{equation*}
This implies $\psi \in \mathcal{D}(H).$ 

Now we prove $\mathcal{D}(H) \subset H^1(\mathbb{R}^d).$ Let $\psi \in \mathcal{D}(H)$. By Lemma $\ref{realizaiton}$ there exists  $\{\psi_n \} \subset C^{\infty}_{0}(\mathbb{R}^d)$ such that 
\begin{equation*} 
\|H \psi_n - H \psi\| + \|\psi_n - \psi\| \rightarrow 0 \ \mathrm{as}\ n \rightarrow \infty.
\end{equation*}
There also exists some $\gamma >0$ such that $ \frac{a}{2} \abs{x}^{2b} + V + \gamma \geq 0 $ on $\mathbb{R}^{d}$.
Hence we have
\begin{equation*}
\frac{1}{2} \|p_{i}(\psi_n-\psi_m)\|^{2}  \leq \langle \psi_n-\psi_m, (H+\gamma) (\psi_n -\psi_m) \rangle  \rightarrow 0 \ \mathrm{as}\ n,m \rightarrow \infty.
\end{equation*}
This implies $\mathcal{D}(H) \subset H^1(\mathbb{R}^d).$
\end{proof}
\subsection{Unitary group and generator}
Let
\begin{equation*}
y :\mathbb{R} \times \mathbb{R}^d \rightarrow \mathbb{R}^d, (t,x) \mapsto y(t,x) = \exp(t \nabla r)(x),
\end{equation*}
be the maximal flow generated by the gradient vector field $\nabla r$. Note that by definition it satisfies
\begin{equation*}
\partial_{t} y^{i}(t,x) = (\nabla r)^{i}(y(t,x)),\ y(0,x) = x.
\end{equation*}
We define $T(t): \mathcal{H} \rightarrow \mathcal{H}, t \in \mathbb{R}$, by
\begin{equation*}
(T(t)\psi)(x) = J(t,x)^{1/2} \psi(y(t,x))
\end{equation*}
where  $J(t,\cdot)$ is the Jacobian of the mapping  $y(t,\cdot): \mathbb{R}^d \rightarrow \mathbb{R}^d.$ By definition $T(t), t \in \mathbb{R}$ forms a strongly continuous one-parameter unitary group. Hence by the Stone theorem the generator $A$ of group $T(t), t \in \mathbb{R}$ is self-adjoint on $\mathcal{H}$. It is easy to verify that $C^{\infty}_0(\mathbb{R}^d) \subset \mathcal{D}(A)$, and that $T(t)$ preserves $C^{\infty}_0(\mathbb{R}^d)$. Hence by \cite[Theorem X.49]{RS} the space $C^{\infty}_0(\mathbb{R}^d)$ is a core for $A$. Therefore, the following facts hold:
\begin{equation*}
\mathcal{D}(A) = \left\{\psi \in \mathcal{H} \ \middle|\  \left(p_{r}-\frac{i}{2}\Delta r\right) \psi \in \mathcal{H}  \right \},
\end{equation*}
\begin{equation*}
A\psi = \left(p_{r}-\frac{i}{2}\Delta r \right)\psi\  (\psi \in \mathcal{D}(A));\ p_{r} = -i \partial_{r}.
\end{equation*}
By Lemma \ref{domain} we also have $\mathcal{D}(H) \subset \mathcal{D}(A)$. Next, we consider describing $H$ using $A$. 
\begin{Lem}
\label{operator1}
One has a decompotision on $C^{\infty}_{0}(\mathbb{R}^d)$
\begin{equation*}
H = \frac{1}{2}A^2+\frac{1}{2}L+q+ \frac{a}{2} \abs{x}^{2b} + V \ \mathrm{with}\  q = \frac{1}{8}(\Delta r)^2 + \frac{1}{4}\partial_{r}(\Delta r).
\end{equation*}
\end{Lem}
\begin{proof}
First, we calculate $\frac{1}{2}A^2$ as
\begin{align*}
\frac{1}{2} A^2 &=\frac{1}{2}\left(p_{r} -  \frac{i}{2}\Delta r \right) \left(p_{r} -  \frac{i}{2}\Delta r \right) \\
                   &= \frac{1}{2}(p_{r})^2 -\frac{i}{4}p_{r}\Delta r -\frac{i}{4}\Delta rp_{r} -\frac{1}{8}(\Delta r)^2 \\
                   & = -\frac{1}{2}(\partial_{i}r) (\partial_{ij}r)\partial_{j} -\frac{1}{2}(\partial_{i}r)(\partial_{j}r)\partial_{ij} -\frac{i}{2}(\Delta r)p_{r} - \frac{1}{4}\partial_{r}(\Delta r) -\frac{1}{8}(\Delta r)^2.
\end{align*}
Next, we calculate $\frac{1}{2}L$ as
\begin{align*}
\frac{1}{2}L &=\frac{1}{2}p_{i}\left( \delta^{ij} - (\partial_{x_i} r)(\partial_{x_j} r) \right)p_j \\
               & = \frac{1}{2}p^{2}_{i}  + \frac{1}{2}(\partial_{i}r) (\partial_{ij}r)\partial_{j}  + \frac{1}{2}(\partial_{i}r)(\partial_{j}r)\partial_{ij}+ \frac{i}{2}(\Delta r)p_{r}.
\end{align*} 
Hence the assertion is verified.
\end{proof}
\subsection{Commutators with weight inside}
Next we introduce a commutator with a weight $\Theta$ inside:
\begin{equation*}
[H,iA]_{\Theta} = i\left(H\Theta A- A\Theta H\right).
\end{equation*}
We assume a weight $\Theta = \Theta(r)$ satisfies the following properties:
\begin{itemize}
\item $\Theta$ is a smooth function with compact support,
\item $r \geq 2$ on $\mathrm{supp}\ {\Theta}$,
\item $\Theta \geq 0$ on $\mathbb{R}$,
\item $\abs{\Theta^{(k)}} \leq C_{k}$, $k=0,1,2,\ldots$
\end{itemize}
where $\Theta^{(k)}$ denotes the $k$-th derivative of $\Theta$ in $r$.
We first define the quadratic form $[H,iA]_{\Theta}$ on $C^{\infty}_{0}(\mathbb{R}^{d})$, and then extend it to $H^{1}(\mathbb{R}^{d})$ according the following lemma.
\begin{Lem}
\label{[]}
As a form on $C^{\infty}_{0}(\mathbb{R}^{d})$,
\begin{equation*}
\begin{split}
[H,iA]_{\Theta} &= A\Theta^{\prime}A + r^{-1}\Theta L -\frac{1}{4}\Theta^{\prime\prime\prime} -(\partial_{r}q)\Theta-ab r^{2b-1}\Theta  -(\partial_{r}V_{1})\Theta  +V_2 \Theta^{\prime} \\
&-2\mathrm{Im}(V_2 \Theta  p_{r} )+(\Delta r) V_2 \Theta -\mathrm{Re}(\Theta^{\prime}H).
\end{split}
\end{equation*}
Therefore by the Cauchy--Schwarz inequality $[H,iA]_{\Theta}$ extends as a bounded form on $H^{1}(\mathbb{R}^d)$.
\end{Lem}
\begin{proof}
By Lemma $\ref{operator1}$ we obtatin
\begin{align*}
[H,iA]_{\Theta} &= \frac{1}{2}[A^2,iA]_{\Theta} + \frac{1}{2}[L,iA]_{\Theta} + [q,iA]_{\Theta} \\ 
 & \quad + \left[\frac{a}{2}\abs{x}^{2b},iA\right]_{\Theta}  + [V_{1},iA]_{\Theta} + [V_{2},iA]_{\Theta}.
\end{align*}
We calculate commutators in turn. First, by $A\Theta - \Theta A = -i \Theta^{\prime}$, we have 
\begin{equation*}
\frac{1}{2}[A^2,iA]_{\Theta} = \frac{1}{2}A\Theta^{\prime}A.
\end{equation*}
Next, we have
\begin{align*}
\frac{1}{2}[L,iA]_{\Theta} &=\mathrm{Im}(A\Theta L) \\
                               &= \mathrm{Im} \left( \left((p_{r})^{*}+\frac{i}{2} \Delta r \right)\Theta L \right)  \\
                               &= \mathrm{Im}(p_{i} (\partial_{i} r) \Theta p_{j} \ell^{jk}p_{k} ) +\frac{1}{2}\mathrm{Re}((\Delta r) \Theta L) \\
                               &= \mathrm{Im} (p_{j} p_{i} (\partial_{i} r) \Theta \ell^{jk}p_{k}) + \mathrm{Re}(p_{i} (\partial_{ij}r) \Theta \ell^{jk}p_{k}) \\
                               &\quad +\frac{1}{2} (\Delta r) \Theta L ,\\
                               \mathrm{Im} (p_{j} p_{i} (\partial_{i} r) \Theta \ell^{jk}p_{k}) &= p_{j } \left(\mathrm{Im}(p_{i} (\partial_{i} r) \Theta \ell^{jk} \right) p_{k} \\
                                                               &= -\frac{1}{2}p_{j} \partial_{i}((\partial_{i} r) \Theta \ell^{jk}) p_{k} \\
                                                               &= -\frac{1}{2}(\Delta r) \Theta L -\frac{1}{2}\Theta^{\prime}L ,\\
      \mathrm{Re}(p_{i} (\partial_{ij}r) \Theta \ell^{jk}p_{k}) &= \mathrm{Re}(p_{i} r^{-1} \ell^{ki} \Theta p_{k})= r^{-1}\Theta L.
\end{align*}
From this we see that the following holds.
\begin{equation*}
\frac{1}{2}[L,iA]_{\Theta} =\mathrm{Im}(A\Theta L)  = r^{-1}\Theta L -\frac{1}{2}\Theta^{\prime}L .
\end{equation*}
Finally, we have
\begin{align*}
[q,iA]_{\Theta} &= -(\partial_{r}q)\Theta -q\Theta^{\prime}, \\
\left[\frac{a}{2}\abs{x}^{2b},iA\right]_{\Theta} &= -ab r^{2b-1}\Theta -\frac{a}{2}r^{2b}\Theta^{\prime}, \\
[V_{1},iA]_{\Theta} &= -(\partial_{r}V_{1})\Theta - V_{1}\Theta^{\prime} , \\
 [V_{2},iA]_{\Theta} &= -2\mathrm{Im}(V_2 \Theta  p_{r} )+(\Delta r) V_2 \Theta.
\end{align*}
We combine the calculations so far. We have
\begin{equation*}
\begin{split}
[H,iA]_{\Theta} &=A\Theta^{\prime}A + r^{-1}\Theta L -(\partial_{r}q)\Theta-ab r^{2b-1}\Theta  -(\partial_{r}V_{1})\Theta  \\
                   &\quad-\mathrm{Re} \left(\frac{1}{2}A\Theta^{\prime}A+ \frac{1}{2}\Theta^{\prime} L + q \Theta^{\prime} + \frac{a}{2}r^{2b}\Theta^{\prime} + V_1 \Theta^{\prime} \right)  \\
                   &\quad -2\mathrm{Im}(V_2 \Theta  p_{r} )+(\Delta r) V_2 \Theta \\
                   &= A\Theta^{\prime}A + r^{-1}\Theta L -(\partial_{r}q)\Theta-ab r^{2b-1}\Theta  -(\partial_{r}V_{1})\Theta -\mathrm{Re}(\Theta^{\prime}H) \\
                    &\quad-\frac{1}{2}\mathrm{Im}(\Theta^{\prime\prime}A) +V_2 \Theta^{\prime} -2\mathrm{Im}(V_2 \Theta  p_{r} )+(\Delta r) V_2 \Theta \\
                   &=A\Theta^{\prime}A + r^{-1}\Theta L -\frac{1}{4}\Theta^{\prime\prime\prime} -(\partial_{r}q)\Theta-ab r^{2b-1}\Theta  -(\partial_{r}V_{1})\Theta  +V_2 \Theta^{\prime} \\
                    &\quad -2\mathrm{Im}(V_2 \Theta  p_{r} )+(\Delta r) V_2 \Theta -\mathrm{Re}(\Theta^{\prime}H).
\end{split}
\end{equation*}
Hence the assertion is verified.
\end{proof}
In the above argument, we defined the weighted commutator $[H,iA]_{\Theta}$ as a quadratic form on $H^{1}(\mathbb{R}^{d})$ as an extension from $C^{\infty}_{0}(\mathbb{R}^{d})$. On the other hand, throughout the paper, we shall use the notation 
\begin{equation*}
\mathrm{Im} (A \Theta H) = \frac{1}{2i} (A \Theta H - H\Theta A)
\end{equation*}
as a quadratic form defined on $\mathcal{D}(H)$, i.e. for $\psi \in \mathcal{D}(H)$
\begin{equation*}
\langle \mathrm{Im} (A \Theta H) \rangle_{\psi} = \frac{1}{2i}\left(\langle A\psi, \Theta H \psi \rangle - \langle H \psi, \Theta A \psi \rangle \right).
\end{equation*}
Note that by Lemma $\ref{domain}$ the above quadratic form is well-defined. Obviously, the quadratic forms $[H,iA]_{\Theta}$ and $2\mathrm{Im} (A \Theta H)$ coincide on $C^{\infty}_{0}(\mathbb{R}^{d})$, and hence we obtain 
\begin{equation*}
[H,iA]_{\Theta} = 2\mathrm{Im} (A \Theta H)\ \mathrm{on}\  \mathcal{D}(H). 
\end{equation*}
In fact, by Lemma $\ref{realizaiton}$ for any $\psi \in \mathcal{D}(H)$ there exists $\{\psi_n\} \subset  C^{\infty}_{0}(\mathbb{R}^d)$ such that
\begin{equation*}
\| \psi_n - \psi \| + \| H(\psi_n - \psi) \| \rightarrow 0 \ \mathrm{as}\ n \rightarrow \infty.
\end{equation*}
Therefore we obtain 
\begin{equation}
\label{[]1}
\langle [H,iA]_{\Theta} \rangle_ \psi = \lim_{n \rightarrow \infty}  \langle [H,iA]_{\Theta} \rangle_{\psi_n} = \lim_{n \rightarrow \infty} \langle 2\mathrm{Im} (A \Theta H) \rangle_{\psi_n} = \langle 2\mathrm{Im} (A \Theta H) \rangle_{\psi}.
\end{equation}
\section{Proof of main results}
In this section we prove the main results by choosing explicit weights and computing commutators.
\subsection{Proof of Theorem $\ref{th1}$}
The proof of Theorem $\ref{th1}$ consists of two steps. Obviously, Theorem $\ref{th1}$ follows immediately as a combination of the following propositions.
\begin{Prop}
\label{Prop1-1}
Let $\lambda \in \mathbb{R}$. If a function $\phi \in \mathcal{H}_\mathrm{loc}$ satisfies that
\begin{itemize}
\item $(H-\lambda)\phi$ =0 in the distributional sense,
\item there exists an $\alpha > \frac{\sqrt{a}}{1+b}$ such that  $\exp({\alpha \abs {x}^{b+1}}) \phi \in \mathcal{H}$, 
\end{itemize}
then $\exp({\alpha \abs {x}^{b+1}}) \phi \in \mathcal{H}$ for any  $\alpha > \frac{\sqrt{a}}{1+b}$.
\end{Prop}
\begin{Prop}
\label{Prop1-2}
Let $\lambda \in \mathbb{R}$. If a function $\phi \in \mathcal{H}_\mathrm{loc}$ satisfies that
\begin{itemize}
\item $(H-\lambda)\phi$ =0 in the distributional sense,
\item $\exp({\alpha \abs {x}^{b+1}}) \phi \in \mathcal{H}$ for any  $\alpha > \frac{\sqrt{a}}{1+b}$,
\end{itemize}
then $\phi =0$ in $\mathbb{R}^{d}$.
\end{Prop}
Now, using the function $\chi \in C^{\infty}(\mathbb{R}^{d})$ of $\eqref{chi}$, we define $\chi_n, \bar{\chi}_n, \chi_{m,n} \in C^{\infty}(\mathbb{R}^{d})$ for $n>m \geq 1$ by
\begin{equation}
\label{chi1}
\chi_{m}(r) = \chi \left(\frac{r}{2^m} \right),\ \bar{\chi}_n = 1- \chi_n,\ \chi_{m,n} = \bar{\chi}_{m}\chi_{n}.
\end{equation} 
To prove Proposition $\ref{Prop1-1}$ we introduce an explicit weight $\Theta$ with parameters $\alpha$, $\beta$, $R$ and $n>m\geq1$:
\begin{equation*}
\label{Theta1}
\Theta = \Theta^{\alpha, \beta}_{m,n,R} = \chi_{m,n} \mathrm{e}^{\theta}.
\end{equation*}
Here the exponent $\theta$ is given by
\begin{equation*}
\theta = 2\alpha r^{b+1} + 2(\beta - \alpha)r^{b+1} \left(1 +\frac{r^{b+1}}{R}  \right)^{-1};\ \beta > \alpha > \frac{\sqrt{a}}{1+b},\ \beta -\alpha \leq 1, \ R > 0.
\end{equation*}
Set for notational simplicity
\begin{equation*}
\theta_{0} = \left(1 +\frac{r^{b+1}}{R}  \right)^{-1},
\end{equation*}
and then 
\begin{align*}
\theta^{\prime} &= 2\alpha (b+1) r^{b} + 2(\beta - \alpha)(b+1)r^{b}\theta^{-2}_{0}, \\
\theta^{\prime \prime} &= 2\alpha b(b+1) r^{b-1} +2(\beta -\alpha) b(b+1) r^{b-1}\theta^{-2}_{0} \\
                               & \quad -4(\beta -\alpha)(b+1)^{2}r^{2b}R^{-1}\theta^{-3}_{0}. 
\end{align*}
In particular since $R^{-1}\theta^{-1}_{0} \leq r^{-(b+1)}$, we have 
\begin{equation*}
\abs{\theta^{\prime \prime \prime}} \leq C(1+ \alpha) r^{b-2}\theta^{-2}_{0}.
\end{equation*}
\begin{Lem}
\label{Ineq1}
Let $\lambda \in \mathbb{R}$, and fix any $\alpha_{0} > \frac{\sqrt{a}}{1+b}$. Then there exist $c, C >0$,  $n_{0} \geq 1$, $\beta > \alpha_{0}$,\ $\alpha_{0} > \tilde{\alpha} $ such that for any $n > m \geq n_{0} $,\ $R>0$,\ $\tilde{\alpha} < \alpha < \alpha_{0}$, 
\begin{equation}
\label{ineq1}
\mathrm{Im}(A\Theta(H-\lambda)) \geq cr^{2b-1}\Theta -C \left(\chi^{2}_{m-1,m+1}+ \chi^{2}_{n-1,n+1}\right)r^{2b-1}\mathrm{e}^{\theta} + \mathrm{Re}(\gamma(H -\lambda))
\end{equation}
as forms on $\mathcal{D}(H)$, where  $\gamma = \gamma_{m,n,R}$ is a function satisfying
\begin{equation*}
\mathrm{supp} \ \gamma \subset \mathrm{supp}\ \chi_{m,n},\ \abs{\gamma} \leq C_{m,n}\mathrm{e}^{\theta}.
\end{equation*}
\end{Lem}
\begin{proof}
Let $\lambda \in \mathbb{R}$ and fix any $\alpha_{0} > \frac{\sqrt{a}}{1+b}$. To be rigorous for the moment all the estimates below are uniform $\frac{\sqrt{a}}{1+b}< \alpha < \alpha_{0}  < \beta ,\ \beta -\alpha \leq 1$, $n >m \geq 1$ and $R > 0$ with constants $C_{\ast}>0$ being independent of them. Then in the last step, we shall restrict ranges of these parameters to obtain assertion. By Lemma $\ref{[]}$, $\eqref{[]1}$, we have 
\begin{equation}
\begin{split}
\label{ineq2}
\mathrm{Im}(A\Theta(H- \lambda)) &= \frac{1}{2}A\theta^{\prime}\Theta A + \frac{1}{2}r^{-1}\Theta L -\frac{1}{8}\Theta^{\prime\prime\prime}+ \frac{1}{2}A\chi^{\prime}_{m,n}\mathrm{e}^{\theta}A -\frac{1}{2}(\partial_{r}q)\Theta \\  
                                   &\quad -\frac{ab}{2} r^{2b-1}\Theta -\frac{1}{2}(\partial_{r}V_{1})\Theta  +\frac{1}{2}V_2\Theta^{\prime}-\mathrm{Im}(V_2 \Theta  p_{r} ) \\ 
                                   & \quad+\frac{1}{2}(\Delta r) V_2 \Theta -\frac{1}{2}\mathrm{Re}(\Theta^{\prime}(H-\lambda)).
\end{split}
\end{equation}
We introduce for simplicity
\begin{equation*}
\begin{split}
Q &= \Bigl((1+\alpha) \chi_{m,n} r^{\max\{2b-1-\mu,b-2,-1\}} + (1+ \alpha^{2}) \abs{\chi^{\prime}_{m,n}} r^{2b} + (1+ \alpha^{2}) \abs{\chi^{\prime \prime}_{m,n}} r^{2b+1} \\ 
&\quad+\abs{\chi^{\prime \prime \prime}_{m,n}}\Bigr)\mathrm{e}^{\theta} + p_{i}\Bigl(\chi_{m,n} + \abs{\chi^{\prime}_{m,n}}\Bigr)\mathrm{e}^{\theta}p_{i}.
\end{split}
\end{equation*}
Let us compute the terms on the right-hand side of $\eqref{ineq2}$ First, by the Cauchy--Schwarz inequality we can estimate 
\begin{equation*}
\frac{1}{2}A\chi^{\prime}_{m,n}\mathrm{e}^{\theta}A-\mathrm{Im}(V_2 \Theta  p_{r} ) \geq -C_{1}Q.
\end{equation*}
Next, computing others , we have
\begin{gather*}
 -\frac{1}{8}\Theta^{\prime\prime\prime} \geq -\frac{1}{8} (\theta^{\prime})^{3}\Theta -\frac{3}{8}\theta^{\prime}\theta^{\prime \prime} \Theta -C_{2}Q, \\
 -\frac{1}{2}(\partial_{r}q)\Theta-\frac{1}{2}(\partial_{r}V_{1})\Theta+\frac{1}{2}V_2\Theta^{\prime}+\frac{1}{2}(\Delta r) V_2 \Theta \geq -C_{3}Q,
\end{gather*}
\begin{align*}
\frac{1}{2}A\theta^{\prime}\Theta A + \frac{1}{2}r^{-1}\Theta L &=\frac{1}{2}Ar^{-1}\Theta A + \frac{1}{2}r^{-1}\Theta L + \frac{1}{2}A\left(\theta^{\prime} - r^{-1}\right) \Theta A \\
                                                                                 &= \frac{1}{2}\mathrm{Re}(r^{-1}\Theta(A^{2}+L)) + \frac{1}{2}\mathrm{Re}(p_{r}(r^{-1}\Theta)A) \\ 
                                                                                 &\quad + \frac{1}{2}A\left(\theta^{\prime} - r^{-1}\right) \Theta A \\
                                                                                 & \geq \mathrm{Re}(r^{-1}\Theta(H- \lambda)) -\frac{a}{2}r^{2b-1}\Theta+ \frac{1}{4}r^{-1}(\theta^{\prime})^{2}\Theta \\
                                                                                 &\quad+ \frac{1}{2}A\left(\theta^{\prime} - r^{-1}\right) \Theta A -C_{4}Q,                                                                              
\end{align*} 
\begin{align*}
\frac{1}{2}A\left(\theta^{\prime} - r^{-1}\right) \Theta A  &\geq \frac{1}{2}\left(A+\frac{i\theta^{\prime}}{2} \right)\left(\theta^{\prime} - r^{-1}\right) \Theta \left(A -\frac{i\theta^{\prime}}{2}\right) \\
&\quad+ \frac{1}{8} (\theta^{\prime})^{3}\Theta -\frac{1}{8}r^{-1}(\theta^{\prime})^{2}\Theta +\frac{1}{2}\theta^{\prime}\theta^{\prime \prime} \Theta -C_{5}Q. 
\end{align*}
We combine the calculations so far. we have 
\begin{align*}
\mathrm{Im}(A\Theta(H- \lambda)) &\geq \frac{1}{8}\theta^{\prime}\theta^{\prime \prime } \Theta + \frac{1}{8}r^{-1}(\theta^{\prime})^{2} \Theta -\frac{a(1+b)}{2}r^{2b-1} \Theta \\
                                   &\quad + \frac{1}{2}\left(A+\frac{i\theta^{\prime}}{2} \right)\left(\theta^{\prime} - r^{-1}\right) \Theta \left(A -\frac{i\theta^{\prime}}{2}\right) \\ 
                                   &\quad+\mathrm{Re} \left(\left(-\frac{1}{2}\Theta^{\prime} + r^{-1}\Theta \right) \left(H-\lambda \right) \right) -C_{6}Q.
\end{align*}
We bound the remainder operator $Q$ as
\begin{align*}
Q &\leq C_{7}(1+\alpha)  r^{\max\{2b-1-\mu,b-2,-1\}} \Theta \\
   &\quad+ C_{7}(1+\alpha^{2})\left(\chi^{2}_{m-1,m+1} + \chi^{2}_{n-1,n+1} \right) r^{2b-1} \mathrm{e}^{\theta} \\
   & \quad+2\mathrm{Re}\left( \left(\chi_{m,n}+ \abs{\chi^{\prime}_{m,n}} \right)\mathrm{e}^{\theta}\left(H -\lambda \right) \right).
\end{align*} 
Hence we obtain
\begin{align*}
\mathrm{Im}(A\Theta(H- \lambda)) &\geq \Bigl(\frac{1}{8}\theta^{\prime}\theta^{\prime \prime }  + \frac{1}{8}r^{-1}(\theta^{\prime})^{2} -\frac{a(b+1)}{2}r^{2b-1}  \\
                                   &\quad - C_{8}(1+\alpha)r^{\max\{2b-1-\mu,b-2,-1\}} \Bigr) \Theta \\
                                   &\quad + \frac{1}{2}\left(A+\frac{i\theta^{\prime}}{2} \right)\left(\theta^{\prime} - r^{-1}\right) \Theta \left(A -\frac{i\theta^{\prime}}{2}\right) \\
                                   &\quad-C_{8} (1+\alpha^{2})\left(\chi^{2}_{m-1,m+1} + \chi^{2}_{n-1,n+1} \right) r^{2b-1} \mathrm{e}^{\theta} \\ 
                                   &\quad+ \mathrm{Re} (\gamma \left(H-\lambda \right)) ,
\end{align*}
where 
\begin{equation*}
\gamma =-\frac{1}{2}\Theta^{\prime} + r^{-1}\Theta -2C_{6}\Theta- 2C_{6}\abs{\chi^{\prime}_{m,n}} \mathrm{e}^{\theta}.
\end{equation*}
Furthermore we calculate the first term as
\begin{align*}
\frac{1}{8}&\theta^{\prime}\theta^{\prime \prime }  + \frac{1}{8}r^{-1}(\theta^{\prime})^{2} -\frac{a(b+1)}{2}r^{2b-1}  - C_{8}(1+\alpha)r^{\max\{2b-1-\mu,b-2,-1\}}  \\
&= \frac{(b+1)}{2}(\alpha^{2}(b+1)^2 - a)r^{2b-1} + \alpha(\beta -\alpha )(b+1)^3r^{2b-1} \theta^{-2}_{0} \\
&\quad +\frac{(\beta -\alpha)^2}{2}(b+1)^3 r^{2b-1}\theta^{-4} -\alpha(\beta -\alpha)(b+1)^3r^{3b}R^{-1}\theta^{-3}_{0} \\
&\quad- (\beta - \alpha)^{2}(b+1)^3r^{3b}R^{-1}\theta^{-5}_{0}- C_{8}(1+\alpha)r^{\max\{2b-1-\mu,b-2,-1\}}.
\end{align*}
From this there exists  $\tilde{\alpha}, \beta$, $n_{0}>0$, $c > 0$ such that for any $\tilde{\alpha} < \alpha < \alpha_{0},\ n>m\geq n_{0},\ R>0$,
\begin{align*}
 \Bigl(\frac{1}{8}\theta^{\prime}\theta^{\prime \prime }  &+ \frac{1}{8}r^{-1}(\theta^{\prime})^{2} -\frac{a(b+1)}{2}r^{2b-1}  \\
                                   &\quad - C_{8}(1+\alpha)r^{\max\{2b-1-\mu,b-2,-1\}} \Bigr) \Theta \geq cr^{2b-1}\Theta.
\end{align*}
By retaking $n_{0}$ larger, if necessary, the third term is non-negative for any $n > m \geq n_{0}$ and $R>0$. Hence the desired estimate follows. 
\end{proof}
\begin{proof}[Proof of Proposition $\ref{Prop1-1}$]
Let $\lambda \in \mathbb{R}$, $\phi \in \mathcal{H}_\mathrm{loc}$ be as in the assertion, and set 
\begin{equation*}
\alpha_{0} = \sup \left\{\alpha \geq \frac{\sqrt{a}}{1+b}\ \middle| \ \exp({\alpha \abs {x}^{b+1}}) \phi \in \mathcal{H} \right \}.
\end{equation*}
By the assumption, we have $\alpha_{0} > \frac{\sqrt{a}}{1+b} $. Assume $\alpha_{0}< \infty$, and we choose $\beta$, $\tilde{\alpha} $ and $n_{0} \geq 1$ as in Lemma $\ref{Ineq1}$.
For any function $\phi$ obeying the assumptions of Proposition $\ref{Prop1-1}$ 
we have $\chi_{m,n} \phi \in H^{2}_{\mathrm{comp}}(\mathbb{R}^d) \subset \mathcal{D}(H) $ for all $n>m \geq 1$.
Note that we may assume $n_{0} \geq 3$, so that for all $n>m \geq n_{0}$
\begin{equation*}
\chi_{m-2,n+2}\phi \in \mathcal{D}(H).
\end{equation*}
We let $ \alpha \in (\tilde{\alpha},\alpha_{0})$. With these parameters fixed evaluate the form inequality from Lemma $\ref{Ineq1}$ on the state $\chi_{m-2,n+2}\phi \in \mathcal{D}(H)$.
Then for any $n>m\geq n_{0}$ and $R>0$
\begin{equation*}
\|(r^{2b-1}\Theta )^{1/2} \phi \|^2 \leq C_{m} \|\chi_{m-1,m+1} \phi \|^2 + C_{R} \| \chi_{n-1,n+1} r^{b- 1/2}  \exp(\alpha r^{b+1})  \phi \|^2.
\end{equation*}
Here we let $\alpha_{1} \in (\alpha,\alpha_{0})$. Then we have 
\begin{align*}
 &\| \chi_{n-1,n+1} r^{b- 1/2} \exp(\alpha r^{b+1} ) \phi \|^2 \\
 &\leq \sup \abs{r^{2b-1 } \exp({2(\alpha -\alpha_{1})r^{b+1} )}}\  \|\chi_{n-1,n+1} \exp({\alpha_{1}r^{b+1}} )\phi \|^2.
\end{align*} 
From this the above second term vanishes as $n \rightarrow \infty$, and consequently by Lebesgue's monotone convergence theorem
\begin{equation*}
\| (r^{2b-1} \bar{\chi}_{m}  \mathrm{e}^{\theta})^{1/2}\phi \|^2 \leq C_{m}\|\chi_{m-1,m+1} \phi \|^2.
\end{equation*}
Next, we let $R \rightarrow \infty$. Again by Lebesgue's monotone convergence theorem it follows that
\begin{equation*}
r^{b-1/2} \bar{\chi}_{m}^{1/2} \exp({\beta r^{b+1}}) \phi \in \mathcal{H}.
\end{equation*}
Thus $ \exp({\beta r^{b+1}}) \phi \in \mathcal{H}$, but this is a contradiction, since $\beta > \alpha_{0}$. Hence we have $\alpha_{0} = \infty$. This implies the assertion.
\end{proof}
To prove Proposition $\ref{Prop1-2}$ we choose a weight 
\begin{equation*}
\Theta = \Theta^{\alpha}_{m,n} = \chi_{m,n} \mathrm{e}^{\theta};\ n > m\geq 1.
\end{equation*}
Here the exponent $\theta$ is given by
\begin{equation*}
\theta = \theta^{\alpha} = 2\alpha r^{b+1};\  \alpha > \frac{\sqrt{a}}{1+b}.
\end{equation*}
Using this weight, we show the following Lemma.
\begin{Lem}
\label{Ineq1-2}
Let $\lambda \in \mathbb{R}$ and $\alpha_{0} > \frac{\sqrt{a}}{1+b} $. Then there exist $c,C>0$ and $n_{0} \geq 1$ such that for any $\alpha >\alpha_{0}$ and $n>m\geq n_{0}$, as quadratic forms on $\mathcal{D}(H)$,
\begin{align*}
\mathrm{Im}(A\Theta(H-\lambda)) &\geq c\alpha^2r^{2b-1}\Theta -C(1+\alpha^2) \left(\chi^{2}_{m-1,m+1}+ \chi^{2}_{n-1,n+1}\right)r^{2b-1}\mathrm{e}^{\theta} \\ 
                                              &\quad + \mathrm{Re}(\gamma(H -\lambda)),
\end{align*}
where $\gamma = \gamma_{m,n,\alpha}$ is a certain function satisfying 
\begin{equation*}
\mathrm{supp}\ \gamma \subset \mathrm{supp}\ \chi_{m,n},\ \abs{\gamma} \leq C_{m,n,\alpha}\mathrm{e}^{\theta}.
\end{equation*}
\begin{proof}
We can prove it similarly to Lemma $\ref{Ineq1}$. Fix any $\lambda \in \mathbb{R}$ and $\alpha_{0} > \frac{\sqrt{a}}{1+b} $. Then, as with the arguments of the proof of Lemma $\ref{Ineq1}$, 
we can estimate for any $\alpha > \alpha$ and $n>m\geq 1$ as
\begin{align*}
\mathrm{Im}(A\Theta(H- \lambda)) &\geq \Bigl(\frac{1}{8}\theta^{\prime}\theta^{\prime \prime }  + \frac{1}{8}r^{-1}(\theta^{\prime})^{2} -\frac{a(b+1)}{2}r^{2b-1}  \\
                                   & \quad- C_{1}(1+\alpha)r^{\max\{2b-1-\mu,b-2,-1\}} \Bigr) \Theta \\
                                   &\quad + \frac{1}{2}\left(A+\frac{i\theta^{\prime}}{2} \right)\left(\theta^{\prime} - r^{-1}\right) \Theta \left(A -\frac{i\theta^{\prime}}{2}\right) \\
                                   &\quad-C_{1} (1+\alpha^{2})\left(\chi^{2}_{m-1,m+1} + \chi^{2}_{n-1,n+1} \right) r^{2b-1} \mathrm{e}^{\theta} \\
                                   &\quad+ \mathrm{Re} (\gamma \left(H-\lambda \right)),
\end{align*}
where 
\begin{equation*}
\begin{split}
Q &= \Bigl((1+\alpha) \chi_{m,n} r^{\max\{2b-1-\mu,b-2,-1\}} + (1+ \alpha^{2}) \abs{\chi^{\prime}_{m,n}} r^{2b} + (1+ \alpha^{2}) \abs{\chi^{\prime \prime}_{m,n}} r^{2b+1} \\ 
&\quad +\abs{\chi^{\prime \prime \prime}_{m,n}} \Bigr)\mathrm{e}^{\theta}  + p_{i}\Bigl(\chi_{m,n} + \abs{\chi^{\prime}_{m,n}}\Bigr)\mathrm{e}^{\theta}p_{i}, \\
\gamma &=-\frac{1}{2}\Theta^{\prime} + r^{-1}\Theta -2C_{1}\Theta- 2C_{1}\abs{\chi^{\prime}_{m,n}} \mathrm{e}^{\theta}.
\end{split}
\end{equation*}
Here a constant $C_1$ is independent of $n,m, \alpha$. Next, we calculate the first term.
\begin{align*}
&\frac{1}{8}\theta^{\prime}\theta^{\prime \prime }  + \frac{1}{8}r^{-1}(\theta^{\prime})^{2} -\frac{a(b+1)}{2}r^{2b-1}  - C_{8}(1+\alpha)r^{\max\{2b-1-\mu,b-2,-1\}}  \\
&\quad =\frac{(b+1)}{2}((b+1)^2\alpha^{2} - a)r^{2b-1} - C_{8}(1+\alpha)r^{\max\{2b-1-\mu,b-2,-1\}}.
\end{align*}
There we choose sufficiently large $n_{0} \geq 1$. Consequently we can easily verify the assertion. Hence we are done.
\end{proof}
\begin{proof}[Proof of Proposition $\ref{Prop1-2}$]
Let $\lambda \in \mathbb{R}$, $\phi \in \mathcal{H}_{\mathrm{loc}}$ be as in the assertion. Fix any $\alpha_{0} > \frac{\sqrt{a}}{1+b}$, and choose $n_{0} \geq 1$ as in the Lemma $\ref{Ineq1-2}$. we may assume that $n_{0} \geq 3$, so that for all $n>m\geq n_{0}$
\begin{equation*}
 \chi_{m-2,n+2}\phi \in \mathcal{D}(H).
\end{equation*}
Evaluate the form inequality from Lemma $\ref{Ineq1-2}$ on the state $\chi_{m-2,n+2}\phi \in \mathcal{D}(H)$, and then for any $\alpha >\alpha_{0}$, $n>m\geq n_{0}$
\begin{equation*}
\|(r^{2b-1}\Theta)^{1/2} \phi \|^2 \leq C_{m} \|\chi_{m-1,m+1} \exp({\alpha r^{b+1}}) \phi \|^2 + \| \chi_{n-1,n+1} r^{b- 1/2} \exp({\alpha r^{b+1}} ) \phi \|^2.
\end{equation*}
Here we let $\alpha_{1}\in (\alpha, \infty)$. Then we have 
\begin{align*}
 &\| \chi_{n-1,n+1} r^{b- 1/2} \exp({\alpha r^{b+1}})  \phi \|^2 \\
 &\leq \sup \abs{r^{2b-1 } \exp({2(\alpha -\alpha_{1})r^{b+1}} )} \  \|\chi_{n-1,n+1} \exp({\alpha_{1}r^{b+1}}) \phi \|^2.
\end{align*} 
From this the above second term to the right vanishes as $n \rightarrow \infty$, and hence by Lebesgue's monotone convergence theorem
\begin{equation*}
\| r^{b-1/2} \bar{\chi}_{m}^{1/2}  \exp({\alpha r^{b+1}}) \phi \|^2 \leq C_{m}\|\chi_{m-1,m+1}  \exp(\alpha r^{b+1}) \phi \|^2.
\end{equation*}
Now assume $\bar{\chi}_{m+2} \phi \not \equiv 0$. The left-hand side grows exponentially as $\alpha \rightarrow \infty$ whereas the right-hand side remains bounded.
This is the contradiction. Thus $\bar{\chi}_{m+2} \phi \equiv 0 $. By invoking the unique continuation property for the second elliptic  operator $H$\ (cf. \cite{Wo})
we conclude that $\phi \equiv 0$ globally on $\mathbb{R}^d$.
\end{proof}
\end{Lem}
\subsection{Proof of Theorem $\ref{th2}$}
The proof of Theorem $\ref{th2}$ consists of two steps. Obviously, Theorem $\ref{th2}$ follows immediately as a combination of the following propositions.
\begin{Prop}
\label{Prop2-1}
Let $\lambda \in \mathbb{R}$. If a function $\phi \in \mathcal{H}_\mathrm{loc}$ satisfies that
\begin{itemize}
\item $(H-\lambda)\phi$ =0 in the distributional sense,
\item there exists an $\abs{\alpha} < \frac{\sqrt{a}}{1+b}$ such that  $\exp({\alpha \abs {x}^{b+1}}) \phi \in \mathcal{H}$, 
\end{itemize}
then $\phi \in \mathcal{H}_{b-1/2}$.
\end{Prop}
\begin{Prop}
\label{Prop2-2}
Let $\lambda \in \mathbb{R}$. If a function $\phi \in \mathcal{H}_\mathrm{loc}$ satisfies that
\begin{itemize}
\item $(H-\lambda)\phi$ =0 in the distributional sense,
\item $\phi \in \mathcal{H}_{b-1/2}$
\end{itemize}
then $ \exp({\alpha \abs {x}^{b+1}}) \phi \in \mathcal{H}$ for any $\abs{\alpha} < \frac{\sqrt{a}}{1+b}$.
\end{Prop}
To prove Proposition $\ref{Prop2-1}$ we introduce a weight $\Theta$ using the function of $\eqref{chi1}$.
\begin{equation*}
\Theta = \Theta^{\alpha}_{m,n,R} = \chi_{m,n} \mathrm{e}^{\theta};\ n > m\geq 1 .
\end{equation*}
Here the exponent $\theta$ is given by
\begin{equation*}
\theta = \theta^{\alpha}_{R} = 2\alpha r^{b+1} - 2 \alpha r^{b+1} \left(1 +\frac{r^{b+1}}{R}  \right)^{-1};\   -\frac{\sqrt{a}}{1+b} <\alpha < 0, \ R > 0.
\end{equation*}
Set for notational simplicity
\begin{equation*}
\theta_{0} = \left(1 +\frac{r^{b+1}}{R}  \right)^{-1},
\end{equation*}
and then 
\begin{align*}
\theta^{\prime} &= 2\alpha(b+1) r^{b} - 2\alpha (b+1)r^{b} \theta^{-2}_{0} ,\\
\theta^{\prime \prime} &= 2\alpha b(b+1)r^{b-1}- 2\alpha b(b+1)r^{b-1} \theta^{-2}_{0}+4\alpha(b+1)^{2}R^{-1} r^{2b}\theta^{-3}_{0}.
\end{align*}
In particular since $R^{-1}\theta^{-1}_{0} \leq r^{-(b+1)}$, we have 
\begin{equation*}
\abs{\theta^{\prime \prime \prime}} \leq C \abs{\alpha} r^{b-2}\theta^{-2}_{0}.
\end{equation*}
\begin{Lem}
\label{Ineq2-1}
Let $\lambda \in \mathbb{R}$, and fix any $ -\frac{\sqrt{a}}{1+b} <\alpha < 0$. Then there exist $c, C>0$ and $n_{0} \geq 1$ such that for any $n>m \geq n_{0}$ and $R>0$, as quadratic form on $\mathcal{D}(H)$,
\begin{equation}
\begin{split}
\label{ineq2-1}
\mathrm{Im}(A\Theta(H-\lambda)) &\leq -cr^{2b-1}\Theta +C \left(\chi^{2}_{m-1,m+1}+ \chi^{2}_{n-1,n+1}\right)r^{2b-1}\mathrm{e}^{\theta} \\
&\quad + \mathrm{Re}(\gamma(H -\lambda)),
\end{split}
\end{equation}
where  $\gamma = \gamma_{m,n,R}$ is a certain function satisfying
\begin{equation*}
\mathrm{supp}\ \gamma \subset \mathrm{supp}\ \chi_{m,n},\ \abs{\gamma} \leq C_{m,n}\mathrm{e}^{\theta}.
\end{equation*}
\begin{proof}
Let $\lambda \in \mathbb{R}$ and $-\frac{\sqrt{a}}{1+b} <\alpha < 0$. 
To be rigorous for the moment all the estimates below are uniform $-\frac{\sqrt{a}}{1+b} <\alpha < 0$, $n >m \geq 1$ and $R > 0$ with constants $C_{\ast}>0$ being independent of them.
By calculating as in Lemma $\ref{Ineq1}$ we can estimate for any $n>m\geq 1$, and $R>0$ as
\begin{align*}
\mathrm{Im}(A\Theta(H- \lambda)) &\leq \frac{1}{8}\theta^{\prime}\theta^{\prime \prime } \Theta + \frac{1}{8}r^{-1}(\theta^{\prime})^{2} \Theta -\frac{a(b+1)}{2}r^{2b-1} \Theta \\
                                   &\quad+ \frac{1}{2}\left(A+\frac{i\theta^{\prime}}{2} \right)\left(\theta^{\prime} - r^{-1}\right) \Theta \left(A -\frac{i\theta^{\prime}}{2}\right) \\ 
                                   & \quad+\mathrm{Re} \left(\left(-\frac{1}{2}\Theta^{\prime} + r^{-1}\Theta \right) \left(H-\lambda \right) \right) +C_{1}Q,
\end{align*}
where 
\begin{equation*}
\begin{split}
Q &= \Bigl((1+\abs{\alpha}) \chi_{m,n} r^{\max\{2b-1-\mu,b-2,-1\}} + (1+ \alpha^{2}) \abs{\chi^{\prime}_{m,n}} r^{2b} + (1+ \alpha^{2}) \abs{\chi^{\prime \prime}_{m,n}} r^{2b+1} \\ 
&\quad +\abs{\chi^{\prime \prime \prime}_{m,n}} \Bigr)\mathrm{e}^{\theta} + p_{i}\Bigl(\chi_{m,n} + \abs{\chi^{\prime}_{m,n}}\Bigl)\mathrm{e}^{\theta}p_{i}.
\end{split}
\end{equation*}
Here we bound the remainder $Q$ as
\begin{align*}
Q &\leq C_{2}(1+\abs{\alpha})  r^{\max\{2b-1-\mu,b-2,-1\}} \Theta \\
   &\quad + C_{2}(1+\alpha^{2})\left(\chi^{2}_{m-1,m+1} + \chi^{2}_{n-1,n+1} \right) r^{2b-1} \mathrm{e}^{\theta} \\
   &\quad +2\mathrm{Re}\left( \left(\chi_{m,n}+ \abs{\chi^{\prime}_{m,n}} \right)\mathrm{e}^{\theta}\left(H -\lambda \right) \right).
\end{align*} 
Hence we have 
\begin{align*}
\mathrm{Im}(A\Theta(H- \lambda)) &\leq \Bigl(\frac{1}{8}\theta^{\prime}\theta^{\prime \prime }  + \frac{1}{8}r^{-1}(\theta^{\prime})^{2} -\frac{a(b+1)}{2}r^{2b-1}  \\
                                   &\quad+ C_{3}(1+\abs{\alpha})r^{\max\{2b-1-\mu,b-2,-1\}} \Bigr) \Theta \\
                                   &\quad  + \frac{1}{2}\left(A+\frac{i\theta^{\prime}}{2} \right)\left(\theta^{\prime} - r^{-1}\right) \Theta \left(A -\frac{i\theta^{\prime}}{2}\right) \\
                                   &\quad +C_{3} (1+\alpha^{2})\left(\chi^{2}_{m-1,m+1} + \chi^{2}_{n-1,n+1} \right) r^{2b-1} \mathrm{e}^{\theta} \\
                                   &\quad+ \mathrm{Re} (\gamma \left(H-\lambda \right)),
\end{align*}
where
\begin{equation*}
\gamma =-\frac{1}{2}\Theta^{\prime} + r^{-1}\Theta +2C_{1}\Theta+ 2C_{1}\abs{\chi^{\prime}_{m,n}} \mathrm{e}^{\theta}.
\end{equation*}
For any $n>m$, we have $\theta^{\prime} - r^{-1} \leq 0$. From this the third term is negative for any $n>m\geq1$.
Next, we compute the first term as
\begin{align*}
&\frac{1}{8}\theta^{\prime}\theta^{\prime \prime }  + \frac{1}{8}r^{-1}(\theta^{\prime})^{2} -\frac{a(1+b)}{2}r^{2b-1}  +C_{3}(1+\abs{\alpha})r^{\max\{2b-1-\mu,b-2,-1\}}  \\
&+ C_{3}(1+\abs{\alpha})r^{\max\{2b-1-\mu,b-2,-1\}} \\
=& \frac{(b+1)}{2}\left((b+1)^{2}\alpha^{2} - a  \right)r^{2b-1} - \alpha^{2}(b+1)^3 r^{2b-1}\theta^{-2}_{0}+\frac{\alpha^2(b+1)^{3}}{2}r^{2b-1}\theta^{-4}_{0} \\
&+ 2\alpha^{2}(b+1)^3r^{4b+1}R^{-2} \theta^{-5}_{0} + \alpha^{2}(b+1)^{3}r^{5b+2} R^{-3} \theta^{-5}_{0} \\ 
&+C_{3}(1+\abs{\alpha})r^{\max\{2b-1-\mu,b-2,-1\}} \\
\leq & \frac{(b+1)}{2}\left((b+1)^{2}\alpha^{2} - a  \right)r^{2b-1} - \alpha^{2}(b+1)^2 r^{2b-1}\theta^{-2}_{0}+\alpha^2(b+1)^{2}r^{2b-1}\theta^{-4}_{0} \\
&+ 2\alpha^{2}(b+1)^3r^{4b+1}R^{-2} \theta^{-5}_{0} + \alpha^{2}(b+1)^{3}r^{5b+2} R^{-3} \theta^{-5}_{0} \\
&+ C_{3}(1+\abs{\alpha})r^{\max\{2b-1-\mu,b-2,-1\}}\\
=& \frac{(b+1)}{2}\left((b+1)^{2}\alpha^{2} - a  \right)r^{2b-1} -2\alpha^{2} (b+1)^{3}R^{-1} r^{3b} \theta^{-4}_{0}\\
&-\alpha^{2} (b+1)^{3}R^{-2} r^{4b+1} \theta^{-4}_{0} + 2\alpha^{2}(b+1)^3r^{4b+1}R^{-2} \theta^{-5}_{0} \\
&+ \alpha^{2}(b+1)^{3} R^{-3} r^{5b+2}  \theta^{-5}_{0} + C_{3}(1+\abs{\alpha})r^{\max\{2b-1-\mu,b-2,-1\}}\\
\leq & \frac{(b+1)}{2}\left((b+1)^{2}\alpha^{2} - a  \right)r^{2b-1} + C_{3}(1+\abs{\alpha})r^{\max\{2b-1-\mu,b-2,-1\}}.
\end{align*}
Here the last inequality can be found from $R^{-1}\theta^{-1}_{0} \leq r^{-(b+1)}$. Thus, the assertion is verified.
\end{proof}
\end{Lem}
\begin{proof}[Proof of Proposition $\ref{Prop2-1}$]
From the assumption we may assume that the constant $\alpha$ satisfies $\alpha<0$.
Let $\lambda \in \mathbb{R}$, $\phi \in  \mathcal{H}_{\mathrm{loc}}$ be as in the assertion, and choose $-\frac{\sqrt{a}}{1+b} < \alpha_{1} <\alpha < 0$. Note that we use Lemma $\ref{Ineq2-1}$ for this constant $\alpha_{1}$, not constant $\alpha$.
We let $n_{0}\geq 1$ as in Lemma $\ref{Ineq2-1}$. Note that we may assume $n_{0} \geq 3$, so that for all  $n>m \geq 3$
\begin{equation*}
 \chi_{m-2,n+2}\phi \in \mathcal{D}(H).
\end{equation*}
Evaluate the form inequality from Lemma $\ref{Ineq2-1}$ on the state $\chi_{m-2,n+2}\phi \in \mathcal{D}(H)$, and then for any $n>m\geq n_{0}$, and $R>0$
\begin{equation*}
\|(r^{2b-1}\Theta)^{1/2} \phi \|^2 \leq C_{m} \|\chi_{m-1,m+1} \phi \|^2 + C_{R} \| \chi_{n-1,n+1} r^{b- 1/2} \exp({\alpha_{1} r^{b+1}})  \phi \|^2.
\end{equation*}
Here we have
\begin{align*}
 &\| \chi_{n-1,n+1} r^{b- 1/2} \exp({\alpha_{1} r^{b+1}})  \phi \|^2 \\
 &{}\leq \sup \abs{r^{2b-1 } \exp({2(\alpha_{1} -\alpha)r^{b+1}) }} \  \|\chi_{n-1,n+1} \exp({\alpha r^{b+1}}) \phi \|^2.
\end{align*} 
Hence by Lebesgue's monotone theorem the assertion is verified.
\end{proof}
To prove Proposition $\ref{Prop2-2}$ we introduce a weight $\Theta$ using the function of $\eqref{chi1}$. 
\begin{equation*}
\Theta = \Theta^{\alpha}_{m,n,R} = \chi_{m,n} \mathrm{e}^{\theta};\ n > m\geq 1 .
\end{equation*}
Here the exponent $\theta$ is given by
\begin{equation*}
\theta = \theta^{\alpha}_{R} = 2 \alpha r^{b+1} \left(1 +\frac{r^{b+1}}{R}  \right)^{-1};\ 0 < \alpha < \frac{\sqrt{a}}{1+b}, \ R > 0.
\end{equation*}
Set for notational simplicity
\begin{equation*}
\theta_{0} = \left(1 +\frac{r^{b+1}}{R}  \right)^{-1},
\end{equation*}
and then 
\begin{align*}
\theta^{\prime} &= 2\alpha(b+1)r^{b}\theta^{-2}_{0} ,\\
\theta^{\prime \prime} &= 2\alpha b(b+1)r^{b-1}\theta^{-2}_{0} - -4\alpha (b+1)^{2} R^{-1} r^{2b} \theta^{-3}_{0}.
\end{align*}
In particular since $R^{-1}\theta^{-1}_{0} \leq r^{-(b+1)}$, we have 
\begin{equation*}
\abs{\theta^{\prime \prime \prime}} \leq C \alpha r^{b-2}\theta^{-2}_{0}.
\end{equation*}
\begin{Lem}
\label{Ineq2-2}
Let $\lambda \in \mathbb{R}$, and fix any $0 <\alpha <  \frac{\sqrt{a}}{1+b}$. Then there exist $c, C>0$ and $n_{0} \geq 1$ such that for any $n>m \geq n_{0}$ and $R>0$, as quadratic form on $\mathcal{D}(H)$, 
\begin{equation}
\begin{split}
\label{ineq2-2}
\mathrm{Im}(A\Theta(H-\lambda)) &\leq -cr^{2b-1}\Theta +C \left(\chi^{2}_{m-1,m+1}+ \chi^{2}_{n-1,n+1}\right)r^{2b-1}\mathrm{e}^{\theta} \\ 
&\quad + \mathrm{Re}(\gamma(H -\lambda)),
\end{split}
\end{equation}
where  $\gamma = \gamma_{m,n,R}$ is a certain function satisfying
\begin{equation*}
\mathrm{supp}\ \gamma \subset \mathrm{supp}\ \chi_{m,n},\ \abs{\gamma} \leq C_{m,n}\mathrm{e}^{\theta}.
\end{equation*}
\end{Lem}
\begin{proof}
Let $\lambda \in \mathbb{R}$ and $0 <\alpha <  \frac{\sqrt{a}}{1+b}$. To be rigorous for the moment all the estimates below are uniform $0 <\alpha <  \frac{\sqrt{a}}{1+b}$, $n >m \geq 1$ and $R > 0$ with constants $C_{\ast}>0$ being independent of them.
Now we can estimate for any $n>m\geq 1$, and $R>0$ as 
\begin{align*}
\mathrm{Im}(A\Theta(H- \lambda)) &\leq \frac{1}{2}A\theta^{\prime}\Theta A + \frac{1}{2}r^{-1}\Theta L -\frac{1}{8} (\theta^{\prime})^{3}\Theta -\frac{3}{8}\theta^{\prime}\theta^{\prime \prime} \Theta  \\
                                   &\quad -\frac{ab}{2} r^{2b-1}\Theta -\frac{1}{2}\mathrm{Re}(\Theta^{\prime}(H-\lambda)) +C_{1}Q,
\end{align*}
where
\begin{equation*}
\begin{split}
Q &= \Bigl((1+\alpha) \chi_{m,n} r^{\max\{2b-1-\mu,b-2,-1\}} + (1+ \alpha^{2}) \abs{\chi^{\prime}_{m,n}} r^{2b} \\
&\quad+ (1+ \alpha^{2}) \abs{\chi^{\prime \prime}_{m,n}} r^{2b+1} +\abs{\chi^{\prime \prime \prime}_{m,n}}\Bigl)\mathrm{e}^{\theta} + p_{i}\Bigl(\chi_{m,n} + \abs{\chi^{\prime}_{m,n}}\Bigr)\mathrm{e}^{\theta}p_{i}.
\end{split}
\end{equation*}
Now by $\theta^{\prime} \geq 0$ we have 
\begin{equation*}
\frac{1}{2}Ar^{-1}\Theta A,\ \frac{1}{2} \theta^{\prime}\Theta L \geq 0.
\end{equation*}
Hence we obtain 
\begin{align*}
\frac{1}{2}A\theta^{\prime}\Theta A &\leq 
\frac{1}{2}A\theta^{\prime}\Theta A +\frac{1}{2}\theta^{\prime}\Theta L \\
&\leq \mathrm{Re} \left(\theta^{\prime} \Theta\left(H-\lambda \right) \right)+ (\lambda - q - \frac{a}{2} r^{2b} -V)\theta^{\prime} \Theta \\ 
& \quad +\frac{3}{4}\theta^{\prime}\theta^{\prime\prime}\Theta + \frac{1}{4}(\theta^{\prime})^{3}\Theta +C_{2}Q, \\
\frac{1}{2}r^{-1}\Theta L &\leq \frac{1}{2}Ar^{-1}\Theta A+ \frac{1}{2}r^{-1}\Theta L \\
&\leq \mathrm{Re} \left(r^{-1} \Theta\left(H-\lambda \right) \right)- \frac{a}{2} r^{2b-1}\Theta \\
& \quad +\frac{1}{4}r^{-1}(\theta^{\prime})^{2}\Theta+C_{3}Q.
\end{align*}
Next we bound the remainder $Q$ as 
\begin{align*}
Q &\leq C_{4}(1+\alpha)  r^{\max\{2b-1-\mu,b-2,-1\}} \Theta \\
   &\quad + C_{4}(1+\alpha^{2})\left(\chi^{2}_{m-1,m+1} + \chi^{2}_{n-1,n+1} \right) r^{2b-1} \mathrm{e}^{\theta} \\
   & \quad +2\mathrm{Re}\left( \left(\chi_{m,n}+ \abs{\chi^{\prime}_{m,n}} \right)\mathrm{e}^{\theta}\left(H -\lambda \right) \right).
\end{align*}
Here we combine the calculations so far. we have
\begin{align*}
\mathrm{Im}(A\Theta(H- \lambda)) &\leq \left(\frac{1}{4} r^{-1} \theta^{\prime} + \frac{3}{8} \theta^{\prime \prime } + \frac{1}{8}(\theta^{\prime})^{2} + \lambda -q - \frac{a}{2}r^{2b} +V  \right)\theta^{\prime}\Theta  \\
                                   & \quad -\frac{a(b+1)}{2}r^{2b-1}\Theta +C_{4}(1+\alpha)  r^{\max\{2b-1-\mu,b-2,-1\}} \Theta \\
                                   & \quad + C_{4}(1+\alpha^{2})\left(\chi^{2}_{m-1,m+1} + \chi^{2}_{n-1,n+1} \right) r^{2b-1} \mathrm{e}^{\theta}\\
                                   &\quad +  \mathrm{Re} \left( \gamma \left(H-\lambda \right)\right),
\end{align*}
where
\begin{equation*}
\gamma =-\frac{1}{2}\Theta^{\prime}+ \theta^{\prime} \Theta+r^{-1} \Theta +2(C_{1}+C_{2}+C_{3})\Theta+ 2(C_{1}+C_{2}+C_{3})\abs{\chi^{\prime}_{m,n}} \mathrm{e}^{\theta}.
\end{equation*}
we compute the first term as
\begin{align*}
&\frac{1}{4} r^{-1} \theta^{\prime} + \frac{3}{8} \theta^{\prime \prime } + \frac{1}{8}(\theta^{\prime})^{2} + \lambda -q - \frac{a}{2}r^{2b} +V \\
&\leq \frac{1}{2}\left((b+1)^{2} \alpha^2 - a \right) r^{2b} + C_{5}(1+ \alpha) r^{\max\{b-1, 2b-1-\mu,0\}}.
\end{align*}
Hence the assertion is verified.
\end{proof}
\begin{proof}[Proof of Proposition $\ref{Prop2-2}$]
Let $\lambda \in \mathbb{R}$, $\phi \in  \mathcal{H}_{\mathrm{loc}}$ be as in the assertion, and choose $0 <\alpha <  \frac{\sqrt{a}}{1+b}$. We let $n_{0} \geq 1$ as in Lemma $\ref{Ineq2-2}$. Note that we may assume $n_{0} \geq 3$, so that for all  $n>m \geq 3$
\begin{equation*}
 \chi_{m-2,n+2}\phi \in \mathcal{D}(H).
\end{equation*}
Evaluate the form inequality from Lemma $\ref{Ineq2-2}$ on the state $\chi_{m-2,n+2}\phi \in \mathcal{D}(H)$, and then for any $n>m\geq n_{0}$, and $R>0$
\begin{equation*}
\|(r^{2b-1}\Theta)^{1/2} \phi \|^2 \leq C_{m} \|\chi_{m-1,m+1} \phi \|^2 + C_{R} \| \chi_{n-1,n+1} r^{b- 1/2}  \phi \|^2.
\end{equation*}
Thus by Lebesgue's monotone theorem the assertion is verified.
\end{proof}
\subsection{Proof of Theorem $\ref{th3}$}
To prove Theorem $\ref{th3}$ we introduce a weight $\Theta$ using the function of $\eqref{chi1}$.
\begin{equation*}
\Theta = \Theta^{\alpha, \beta}_{m,n,R} = \chi_{m,n} \mathrm{e}^{\theta};\ n > m\geq 1 .
\end{equation*}
Here the exponent $\theta$ is given by
\begin{equation*}
\theta = 2\alpha r^{b+1} + 2(\beta - \alpha)r^{b+1} \left(1 +\frac{r^{b+1}}{R}  \right)^{-1};\ \alpha< \beta <- \frac{\sqrt{a}}{1+b},\ \beta -\alpha \leq 1, \ R > 0.
\end{equation*}
Set for notational simplicity
\begin{equation*}
\theta_{0} = \left(1 +\frac{r^{b+1}}{R}  \right)^{-1}.
\end{equation*}
Using this weight, we show the following Lemma.
\begin{Lem}
\label{Ineq3-1}
Let $\lambda \in \mathbb{R}$, and fix any $\alpha_{0} <- \frac{\sqrt{a}}{1+b}$. Then there exist $c, C>0$, $n_{0} \geq 1$, $\beta > \alpha_{0}$, $\alpha_{0} > \tilde{\alpha} $ such that for any $n > m \geq n_{0}$, $R>0$, $\tilde{\alpha} < \alpha < \alpha_{0}$, 
\begin{align*}
\label{ineq3-1}
\mathrm{Im}(A\Theta(H-\lambda)) &\leq -cr^{3b}\Theta + \frac{1}{2}\left( r^{-1} -\theta^{\prime } \right) \Theta L \\ \notag
&\quad+C \left(\chi^{2}_{m-1,m+1}+ \chi^{2}_{n-1,n+1}\right)r^{2b-1}\mathrm{e}^{\theta} + \mathrm{Re}(\gamma(H -\lambda)) \notag
\end{align*}
as forms on $\mathcal{D}(H)$, where $\gamma = \gamma_{m,n,R}$ is a function satisfying 
\begin{equation*}
\mathrm{supp}\ \gamma \subset \mathrm{supp}\ \chi_{m,n},\ \abs{\gamma} \leq C_{m,n}\mathrm{e}^{\theta}.
\end{equation*}
\end{Lem}
\begin{proof}
Let $\lambda \in \mathbb{R}$, and fix any $\alpha_{0} <- \frac{\sqrt{a}}{1+b}$. To be rigorous for the moment all the estimates below are uniform $\alpha < \alpha_{0}  < \beta <- \frac{\sqrt{a}}{1+b} ,\ \beta -\alpha \leq 1$, $n >m \geq 1$ and $R > 0$ with constants $C_{\ast}>0$ being independent of them. We can estimate
\begin{align*}
\mathrm{Im}(A\Theta(H- \lambda)) &\leq \frac{1}{2}A\theta^{\prime}\Theta A + \frac{1}{2}r^{-1}\Theta L -\frac{1}{8} (\theta^{\prime})^{3}\Theta -\frac{3}{8}\theta^{\prime}\theta^{\prime \prime} \Theta  \\
                                   &\quad -\frac{ab}{2} r^{2b-1}\Theta -\frac{1}{2}\mathrm{Re}(\Theta^{\prime}(H-\lambda))+C_{1}Q,
\end{align*}
where
\begin{equation*}
\begin{split}
Q &=\Bigl((1+\abs{\alpha}) \chi_{m,n} r^{\max\{2b-1-\mu,b-2,-1\}} + (1+ \alpha^{2}) \abs{\chi^{\prime}_{m,n}} r^{2b} + (1+ \alpha^{2}) \abs{\chi^{\prime \prime}_{m,n}} r^{2b+1} \\
& \quad +\abs{\chi^{\prime \prime \prime}_{m,n}}\Bigr)\mathrm{e}^{\theta} + p_{i}\Bigl(\chi_{m,n} + \abs{\chi^{\prime}_{m,n}}\Bigr)\mathrm{e}^{\theta}p_{i}.
\end{split}
\end{equation*}
We also have 
\begin{align*}
\frac{1}{2}A\theta^{\prime}\Theta A &= \frac{1}{2}A\theta^{\prime}\Theta A + \frac{1}{2}\theta^{\prime}\Theta L  -\frac{1}{2}\theta^{\prime}\Theta L\\
                                              & \leq \left(\lambda - q - \frac{a}{2}r^{2b} -V\right) \theta^{\prime} \Theta +\frac{3}{4}\theta^{\prime}\theta^{\prime \prime} \Theta + \frac{1}{4}(\theta^{\prime})^{3}\Theta \\ 
                                              &\quad + \mathrm{Re}\left(\theta^{\prime} \Theta \left(H - \lambda \right)\right) -\frac{1}{2}\theta^{\prime}\Theta L.
\end{align*}
Next we bound the remainder $Q$ as
\begin{align*}
Q &\leq C_{2}(1+\alpha)  r^{\max\{2b-1-\mu,b-2,-1\}} \Theta + C_{2}(1+\alpha^{2})\left(\chi^{2}_{m-1,m+1} + \chi^{2}_{n-1,n+1} \right) r^{2b-1} \mathrm{e}^{\theta} \\
   &\quad +2\mathrm{Re}\left( \left(\chi_{m,n}+ \abs{\chi^{\prime}_{m,n}} \right)\mathrm{e}^{\theta}\left(H -\lambda \right) \right).
\end{align*} 
We combine the calculations so far as
\begin{align*}
\mathrm{Im}(A\Theta(H- \lambda)) &\leq \left(\left(\lambda - q - \frac{a}{2}r^{2b} -V\right) \theta^{\prime}  +\frac{3}{8}\theta^{\prime}\theta^{\prime \prime}  + \frac{1}{8}(\theta^{\prime})^{3}\right) \Theta  \\ 
& \quad+C_{3}(1+\alpha)  r^{\max\{2b-1-\mu,b-2,-1\}} \Theta + \frac{1}{2}\left( r^{-1} -\theta^{\prime } \right) \Theta L \\
&\quad + C_{3}(1+\alpha^{2})\left(\chi^{2}_{m-1,m+1} + \chi^{2}_{n-1,n+1} \right) r^{2b-1} \mathrm{e}^{\theta}\\
&\quad+ \mathrm{Re} \left(\gamma  \left(H - \lambda \right) \right),
\end{align*}
where
\begin{equation*}
\gamma =-\frac{1}{2}\Theta^{\prime} + \theta^{\prime}\Theta +2C_{1}\Theta+ 2C_{1}\abs{\chi^{\prime}_{m,n}} \mathrm{e}^{\theta}.
\end{equation*}
Furthermore we have
\begin{align*}
 &\left(\left(\lambda - q - \frac{a}{2}r^{2b} -V\right) \theta^{\prime}  +\frac{3}{8}\theta^{\prime}\theta^{\prime \prime}  + \frac{1}{8}(\theta^{\prime})^{3}\right) \\
 &\leq - \frac{a}{2}r^{2b} \theta^{\prime} + \frac{1}{8} (\theta^{\prime})^{3}+C_{5}(1+\abs{\alpha})r^{\max\{b,3b-1-\mu,2b-1\}} \\
 &=\left(b+1\right)\left((b+1)^2\alpha^{3} - a \alpha  \right) r^{3b} + 3\alpha^{2}(b+1)^{3}(\beta - \alpha)r^{3b}\theta^{-2}_{0}\\
 & \quad- a(b+1)(\beta- \alpha) r^{3b}\theta^{-2}_{0} +3a \alpha(\beta- \alpha)^{2}r^{3b} \theta^{-4}_{0} \\
 & \quad + (b+1)^{3}(\beta - \alpha)^{3} r^{3b}\theta^{-6}_{0} +C_{5}(1+\abs{\alpha})r^{\max\{b,3b-1-\mu,2b-1\}}.
\end{align*}
Hence the assertion is verified.
\end{proof}
\begin{proof}[Proof of Theorem $\ref{th3}$]
Let $\lambda \in \mathbb{R}$, $\phi \in \mathcal{H}_{\mathrm{loc}} $ be as in the assertion, and set 
\begin{equation*}
\alpha_{0} = \sup \left\{\alpha \leq - \frac{\sqrt{a}}{1+b}\ \middle| \ \exp({\alpha \abs{x}^{1+b} } )\phi \in \mathcal{H} \right \}.
\end{equation*}
From the assumpton we have $\alpha_{0} > -\infty $. Assume $\alpha_{0} < - \frac{\sqrt{a}}{1+b} $, and we choose $\beta$, $\tilde{\alpha} $ and $n_{0} \geq 1$ as in Lemma $\ref{Ineq3-1}$.
Note that we may assume $n_{0} \geq 3$, so that for all $n>m \geq n_{0}$
\begin{equation*}
\chi_{m-2,n+2}\phi \in \mathcal{D}(H).
\end{equation*}
Evaluate the form inequality from Lemma $\ref{Ineq3-1}$ on the state $\chi_{m-2,n+2}\phi \in \mathcal{D}(H)$, and then for any $n>m\geq n_{0}$, and $R>0$
\begin{equation*}
\|(r^{3b} \Theta)^{1/2} \phi \|^2 \leq C_{m} \|\chi_{m-1,m+1} \phi \|^2 + C_{R} \| \chi_{n-1,n+1} r^{b- 1/2} \exp({\alpha r^{b+1}})  \phi \|^2.
\end{equation*}
Note that we have used the condition on $L$ in the assumption. Now we let $\alpha_{1} \in (\alpha, \alpha_{0})$. Then we have 
\begin{align*}
 &\| \chi_{n-1,n+1} r^{b- 1/2} \exp(\alpha r^{b+1})  \phi \|^2 \\
 &{}\leq \sup \abs{r^{2b-1 } \exp({2(\alpha -\alpha_{1})r^{b+1}})}\  \|\chi_{n-1,n+1} \exp({\alpha_{1}r^{b+1}}) \phi \|^2.
\end{align*} 
From this the above second term vanishes as $n \rightarrow \infty$, and consequently by Lebesgue's monotone convergence theorem 
\begin{equation*}
\| (r^{3b} \bar{\chi}_{m}  \mathrm{e}^{\theta})^{1/2} \phi \|^2 \leq C_{m}\|\chi_{m-1,m+1} \phi \|^2.
\end{equation*}
Next, we let $R \rightarrow \infty$. Again by Lebesgue's monotone convergence theorem it follows that 
\begin{equation*}
r^{b-1/2} \bar{\chi}_{m}^{1/2} \exp({\beta r^{b+1}}) \phi \in \mathcal{H}.
\end{equation*}
Thus $ \exp({\beta r^{b+1}}) \phi \in \mathcal{H}$, but this is a contradiction, since $\beta > \alpha_{0}$. Hence we have $\alpha_{0} \geq - \frac{\sqrt{a}}{1+b}$. This implies the assertion.
\end{proof}
\section*{Acknowledgement}
The author would like to thank his supervisor Kenichi Ito for discussions and encouragement. 

\end{document}